\documentclass[12pt]{iopart}

\usepackage[scale=1.0]{OldStandard}
\usepackage{libertinust1math}
\usepackage{mylatexcommands}
\usepackage{qmdefs}
\usepackage{color}
\usepackage{graphicx}
\usepackage{ifthen}
\usepackage[]{cite}
\usepackage[]{mcite}

\newcommand{\figdir}{.}  

\newboolean{revision}
\setboolean{revision}{false}
\ifthenelse{\boolean{revision}}
{\newcommand{\authorcorr}{\color{red}}    \newcommand{\refereeone}{\color{blue}}     }  
{\newcommand{\authorcorr}{\color{black}}  \newcommand{\refereeone}{\color{black}}      } 

\definecolor{grey}{gray}{0.5}

\newlength{\ftwidth}   
\newlength{\sfsize}   

\begin{document}

\newcommand{\gguide}{{\it Preparing graphics for IOP Publishing journals}}

\title[On Ehrenfest's theorem]{Addendum: Considerations about the incompleteness of the Ehrenfest's theorem in quantum mechanics (2021 Eur. J. Phys. 42 065405)}

\author{Domenico Giordano}
\address{European Space Agency (retired), The Netherlands}  
\ead{dg.esa.retired@gmail.com}

\author{Pierluigi Amodio}
\address{Dipartimento di Matematica, Universit\`a di Bari, Italy}
\ead{pierluigi.amodio@uniba.it}

\vspace{10pt}
\begin{indented}
\item[]\today
\end{indented}

\renewcommand {\REq}  [1] {\mbox{equation~(\ref{#1})}}                        
\renewcommand {\REqq} [1] {\mbox{equations~(\ref{#1})}}                       
\renewcommand {\REqqc}[2] {\mbox{equations~(\ref{#1})#2}}                     
\renewcommand {\Rse}  [1] {section~\ref{#1}}                                  
\renewcommand {\Rfi}  [1] {\mbox{figure~\ref{#1}}}                            

\newcommand{\Reqma}  [1] {\Reql{equation} {#1}{\mbox{\scriptsize  ma}}}
\newcommand{\Reqsma} [1] {\Reql{equations}{#1}{\mbox{\scriptsize  ma}}}

\begin{abstract}  We describe the analytical solution of the eigenvalue problem introduced in our article mentioned in the title and relative to a punctiform electric charge confined in an one-dimensional box in the presence of an electric field.
We also derive and discuss the analytical expressions of the external forces acting on the punctiform charge and associated with the boundaries of the one-dimensional box in the presence of the electric field.  \end{abstract}

%
%
%
%
%

\noindent Accepted for publication in the \textit{European Journal of Physics} on 16 December 2021 \newline
https://iopscience.iop.org/article/10.1088/1361-6404/ac43f3 \newline
This Accepted Manuscript is available for reuse under CC BY-NC-ND licence after the 12 month embargo period provided that all the terms and conditions of the licence are adhered to.


\section{Introduction\label{intro}}
The additional reflections described here complement our considerations{\authorcorr\cite{dg2021ejp}} on the Ehrenfest's theorem
and were inspired by the feedback we received from K. Moulopoulos\,\footnote{Department of Physics, University of Cyprus.} in a private communication.
He and his colleagues also produced interesting contributions \cite{gk2016ijeir,gk2018jopc,km2020arxiv} to the Ehrenfest-theorem theme that, unfortunately and regrettably, escaped the literature survey we carried out when working on our main article \cite{dg2021ejp}; our findings turn out to be in accordance with those exposed in the cited contributions which, by the way, have the valuable feature of not being restricted to spatially one-dimensional cases.
Of course, the reading of this short addendum presupposes familiarity with our main article and we assume it acquired by the reader. 
In the sequel, we will use the same notation adopted therein and will make reference to its equations by subscripting the label `ma' (main article); for example, \Reqma{1} refers to equation~(1) in the main article.

Moulopoulos' comment regards \Reqma{104} which gives the dimensional expression of the external forces due to the confining walls acting on the punctiform electric charge when the electric field is absent.
As duly reported in \cite{sdv2013jop,sdv2013rbef,dg2021ejp}, ter Haar \cite{dth1964} derived the same result in 1964 by following a different approach\,\footnote{More bibliographic information is provided in \cite{dg2021ejp}.} and emphasized its coincidence with the classical-mechanics result.
With due account of \Reqsma{94} and \Reqma{100}, the nondimensional version of \Reqma{104} reads
\begin{equation}\label{ef.dv.c}
   \left( \pd{}{\psik}{\xi} \pd{}{\psikc}{\xi} \right)_{\xi=\pm 1} = 2\,\beta_{k} 
\end{equation}
Both \Reqma{104} and \REq{ef.dv.c} indicate that, for each eigenstate, the external forces depend linearly on the energy eigenvalue when the electric field is absent.
Moulopoulos' question addresses the circumstance in which the electric field is present; in his own words: 
\begin{quote}
  From your later numerical results (the ones that show the several interesting extrema), do I get it correctly that this CLASSICAL relation is no more satisfied for non-zero electric fields?
\end{quote}
He refers to the results shown in the graphs of \mbox{figure~2} of \cite{dg2021ejp} and definitely perceives the appropriate conclusion regarding the ``CLASSICAL relation'' [\Reqma{104}].
Indeed, for a given eigenstate $k$, left and right slopes of the eigenfunction differ when \mbox{$\alpha=10,100$} but the energy eigenvalue is unique; therefore, the linear dependence indicated by \REq{ef.dv.c}, or \Reqma{104}, must break down somehow.
So, what kind of functional dependence is there between external forces and energy eigenvalue in the presence of the electric field?
The quantitative answer to this question requires the analytical solution of the eigenvalue problem defined by \Reqsma{95}\,.
This is indeed possible and we will deal with such a task in \Rse{asev}.
In \Rse{efeef}, we will present and discuss the consequences on the external forces that follow from the analytical solution.
In \Rse{ap}, we will present the analytical proof of \Reqma{105}, which corroborates the numerical verification described in \cite{dg2021ejp}.

\section{Analytical solution of the eigenvalue problem\label{asev}}

For convenience, we rewrite here the \Reqsma{95}
\begin{subequations}\label{evp}
  \begin{equation}\label{evp.se}
    - \pd{2}{\psi}{\xi} - \alpha \xi \psi = \beta \psi
  \end{equation} 
  \vspace{-0.5\baselineskip}
  \begin{equation}\label{evp.bc}
    \psi(-1) = \psi(+1) = 0
  \end{equation}
  \vspace{-0.3\baselineskip}
  \begin{equation}\label{evp.nc}
    \frac{1}{2} \int_{-1}^{+1} \!\!\!\psi^{\ast}\psi \, d\xi = 1
  \end{equation}
\end{subequations}
that define the eigenvalue problem.
We are interested in the situation with the presence of the electric field and, therefore, we assume \mbox{$\alpha\neq0$}.
Then, the independent-variable transformation
\begin{equation}\label{ivt}
  \xi = - \dfrac{\eta}{\alpha^{1/3}} - \dfrac{\beta}{\alpha}
\end{equation}
converts the time-independent Schr\"{o}dinger equation [\REq{evp.se}] into the Airy differential equation
\begin{equation}\label{evp.se.a}
  \pd{2}{\psi}{\eta} - \eta\,\psi = 0
\end{equation} 
whose general solution is a linear combination of the Airy functions
\begin{equation}\label{evp.se.a.gs}
  \psi(\eta) = A\cdot\Ai(\eta) + B\cdot\Bi(\eta) 
\end{equation} 
Next step consists in the imposition of the boundary conditions [\REq{evp.bc}].
The values of the new independent variable $\eta$ at the left/right boundaries \mbox{($\xi=\mp 1$)} are obtained from \REq{ivt} by inversion; they read respectively
\begin{subequations}\label{ivbv}
	\begin{align}
	   \etah = & - \dfrac{\beta-\alpha}{\alpha^{2/3}}         \label{ivbv.l}\\[2ex]
	   \bar{\eta} = & - \dfrac{\beta+\alpha}{\alpha^{2/3}}    \label{ivbv.r}
	\end{align}
\end{subequations}
Then, the imposition of \REq{evp.bc} leads to the algebraic homogeneous system
\begin{equation}\label{bc.c}
  \begin{cases}
      \psi(-1) \rightarrow \psi(\etah) = A\cdot\Ai(\etah) + B\cdot\Bi(\etah) = 0 & \qquad \text{left boundary}  \\[3ex]  
      \psi(+1) \rightarrow \psi(\etab) = A\cdot\Ai(\etab) + B\cdot\Bi(\etab) = 0 & \qquad \text{right boundary}
  \end{cases}
\end{equation}
for the coefficients $A,B$ and its determinant's vanishing leads to the algebraic equation
\begin{equation}\label{evg}
   \Ai(\etah) \cdot \Bi(\etab) - \Ai(\etab) \cdot \Bi(\etah) = 0
\end{equation}
that generates the eigenvalues.
\REqb{evg} requires a numerical solution and we found out that the Newton-Raphson method works very well for that purpose.%
\begin{table}[h]
   \hspace*{\mathindent}
   \resizebox{.98\ftwidth}{!}{\includegraphics*{\figdir/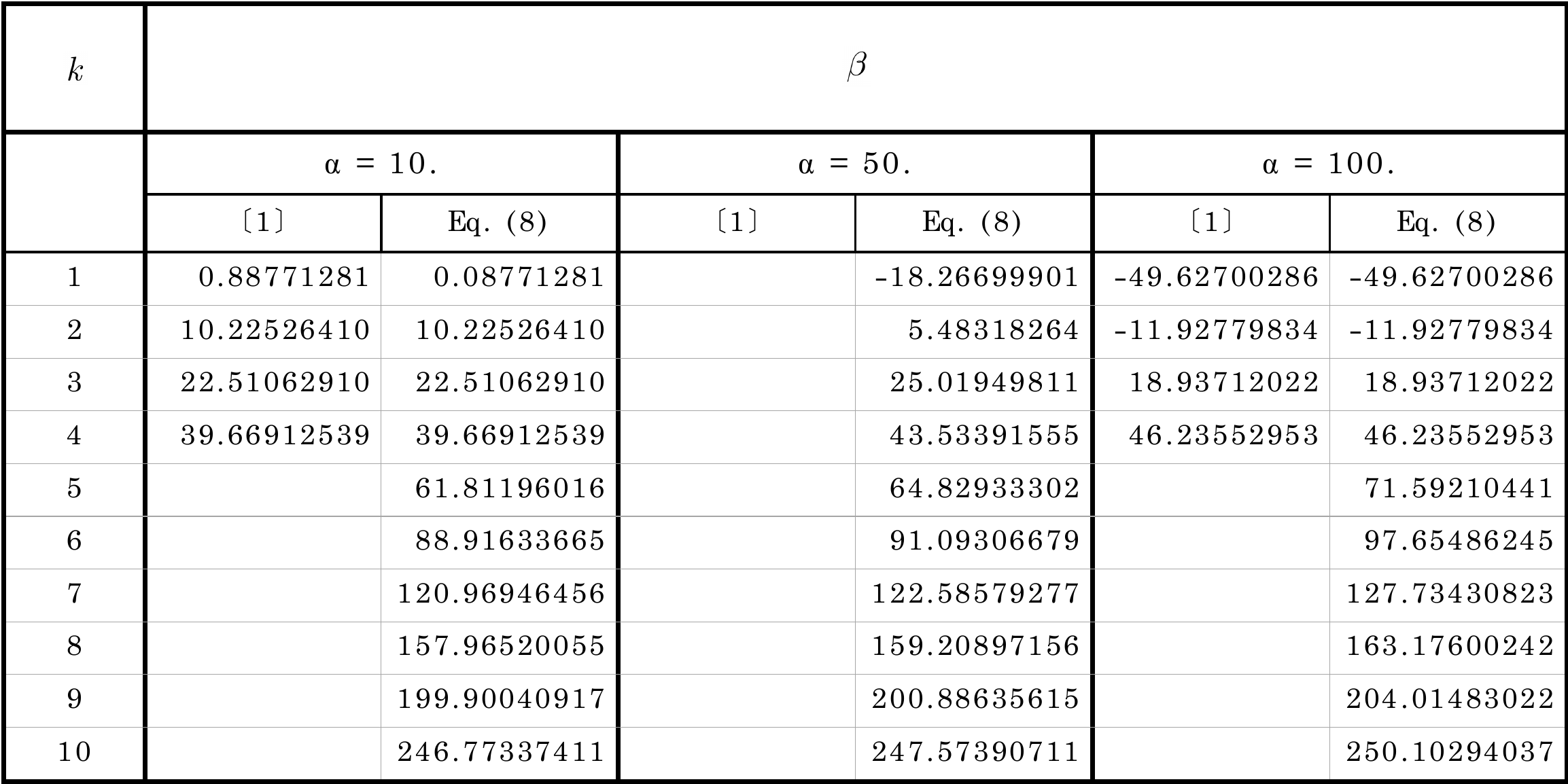}}  
   \caption{Eigenvalues calculated from \REq{evg} and from the numerical algorithm of \cite{dg2021ejp}\label{table}}
\end{table}
\Rta{table} shows the first ten eigenvalues obtained from \REq{evg} for three values of the electric field \mbox{($\alpha=10,50,100$)}; for comparison and validation, we have included also the first four eigenvalues (\mbox{$\alpha=10,100$} only) produced by the finite-difference algorithm described in \cite{dg2021ejp} to solve numerically the complete eigenvalue problem [\REqq{evp}].
With the eigenvalues in hand, the algebraic homogeneous system provides one relation between the coefficients $A,B$; if we choose to express $B$ in terms of $A$\,\footnote{The other way round is obviously equivalent.}
\begin{equation}\label{B}
  B = - A \cdot w
\end{equation}
in which, for brevity, we have set
\begin{equation}\label{w}
   w = \dfrac{\Ai(\etah)}{\Bi(\etah)} = \dfrac{\Ai(\etab)}{\Bi(\etab)}
\end{equation}
then the analytical eigenfunction [\REq{evp.se.a.gs}] turns into the form
\begin{equation}\label{evp.se.a.gs.1}
  \psi(\eta) = A \cdot f(\eta)
\end{equation} 
in which, again for brevity, we have set
\begin{equation}\label{feta}
   f(\eta) = \Ai(\eta) -w \cdot \Bi(\eta)
\end{equation}
The coefficient $A$ is determined by the normalization condition [\REq{evp.nc}].
The latter's exploitation requires a bit of care to rephrase the integral in terms of the new independent variable via \REq{ivt} but the normalization operation is rather elementary and leads to
\begin{equation}\label{A}
   A = \sqrt{\dfrac{2\alpha^{1/3}}{J}}
\end{equation}
with
\begin{equation}\label{J}
  J = J(\alpha) = \int_{\etab}^{\etah} f(\eta)^{2} d\eta
\end{equation}
Obviously, the integral in \REq{J} can be evaluated numerically but there is an analytical shortcut which we will return to and describe in more details in \Rse{efeef}.
The substitution of \REq{A} into \REq{evp.se.a.gs.1} yields the final form
\begin{equation}\label{evp.se.a.gs.2}
  \psi(\eta) = \sqrt{\dfrac{2\alpha^{1/3}}{J}} \cdot f(\eta)
\end{equation} 
of the analytical eigenfunctions.
In \Rfi{aef1-4.wef}, we provide validation evidence by superposing the numerical eigenfunctions calculated in \cite{dg2021ejp} on the analytical eigenfunctions produced by \REq{evp.se.a.gs.2} for the first four eigenstates.
\begin{figure}[h]
   \hspace*{\mathindent}
   \resizebox{\sfsize}{!}{\includegraphics*[trim=10 25 60 60]{\figdir/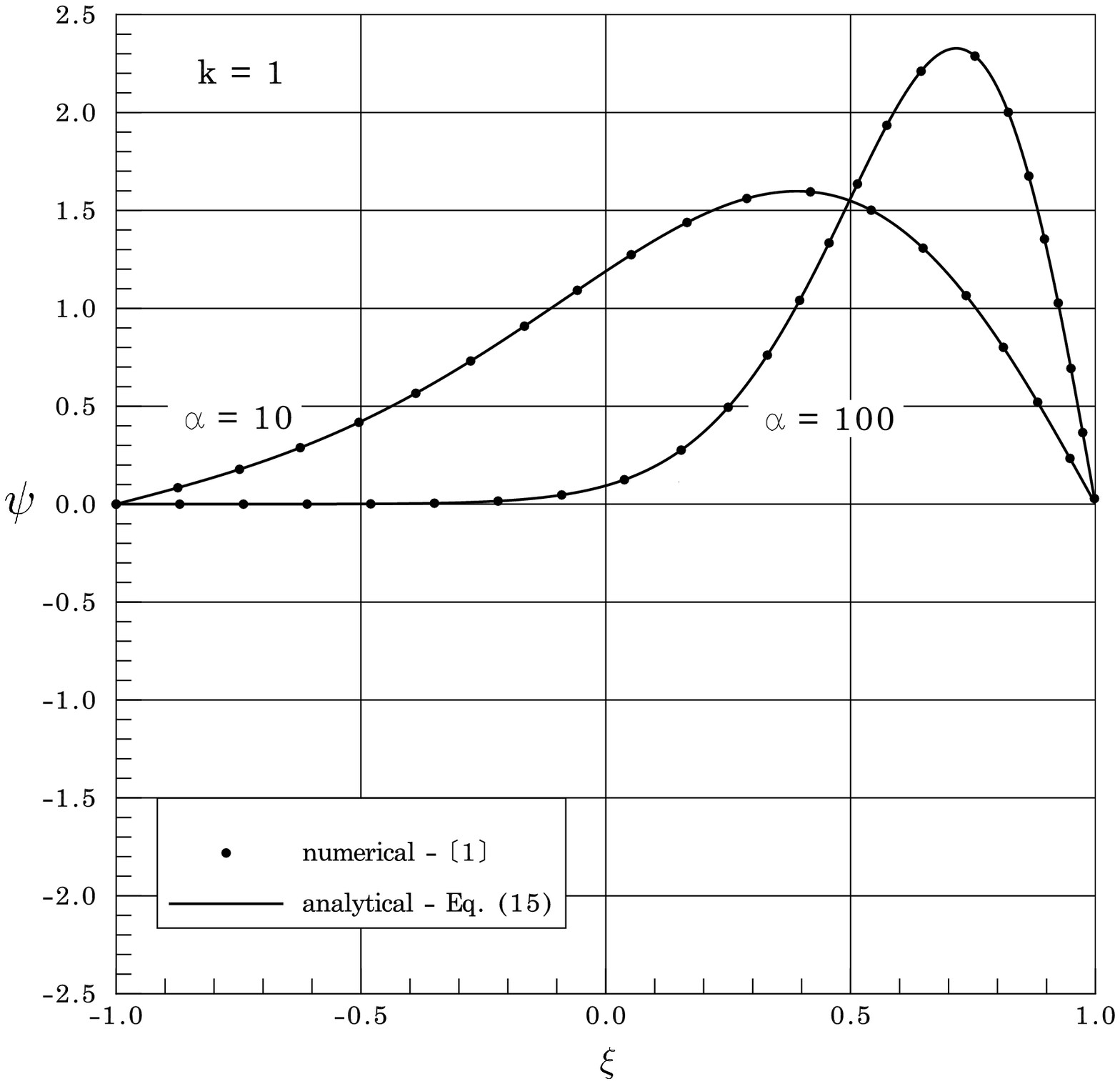}} 
   \resizebox{\sfsize}{!}{\includegraphics*[trim=10 25 60 60]{\figdir/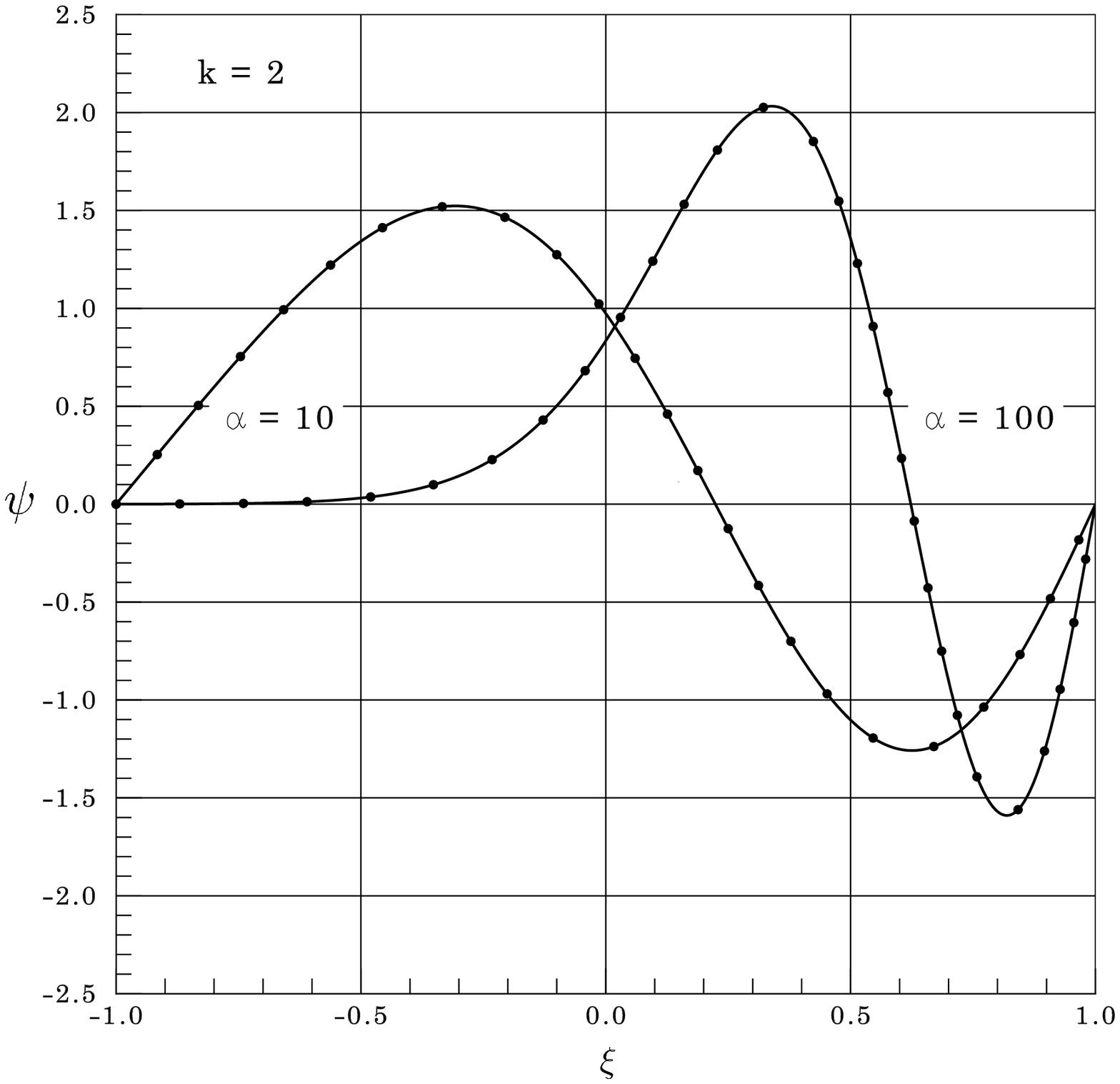}} \\ 
   \hspace*{\mathindent}
   \resizebox{\sfsize}{!}{\includegraphics*[trim=10 25 60 60]{\figdir/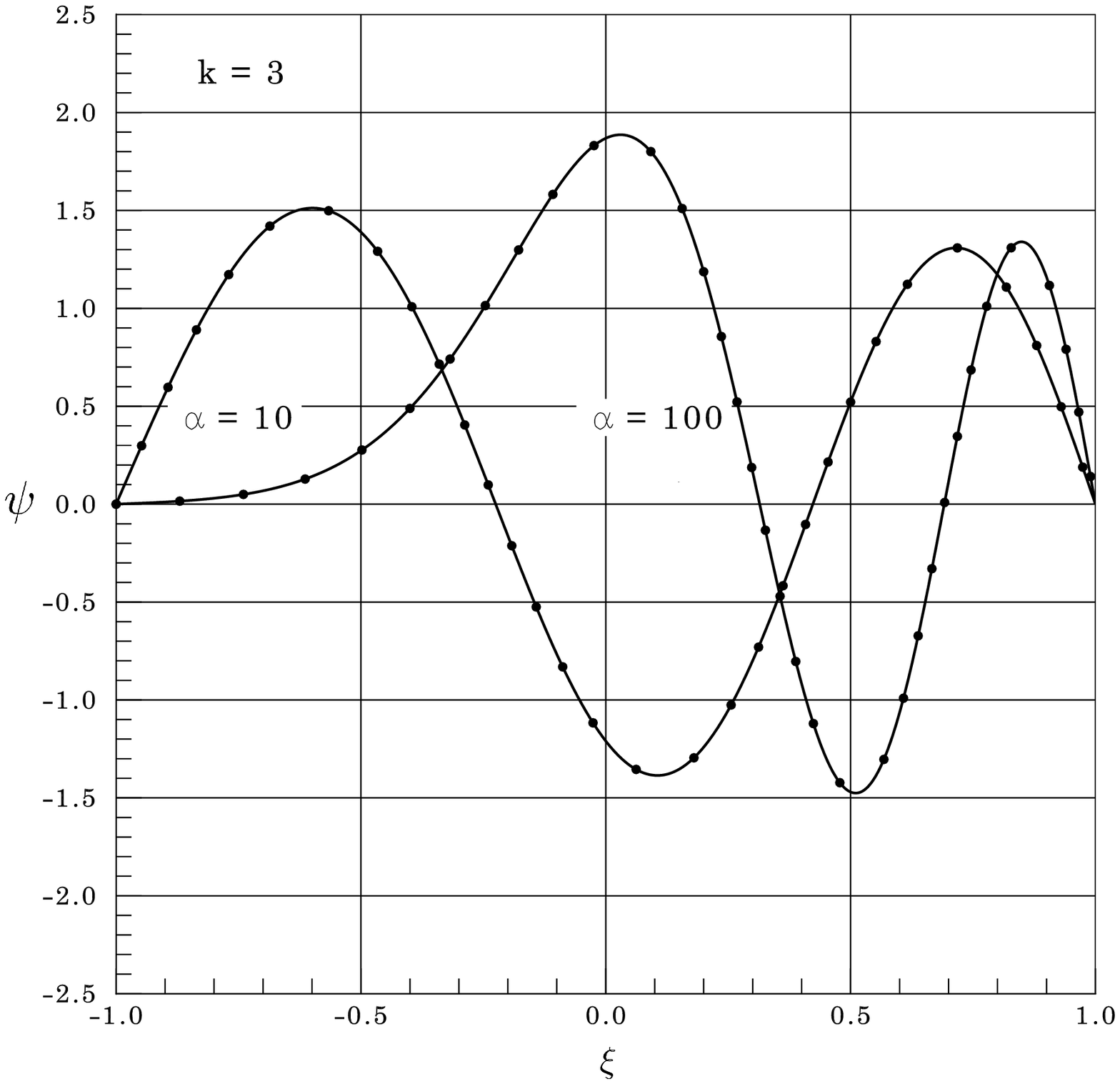}} 
   \resizebox{\sfsize}{!}{\includegraphics*[trim=10 25 60 60]{\figdir/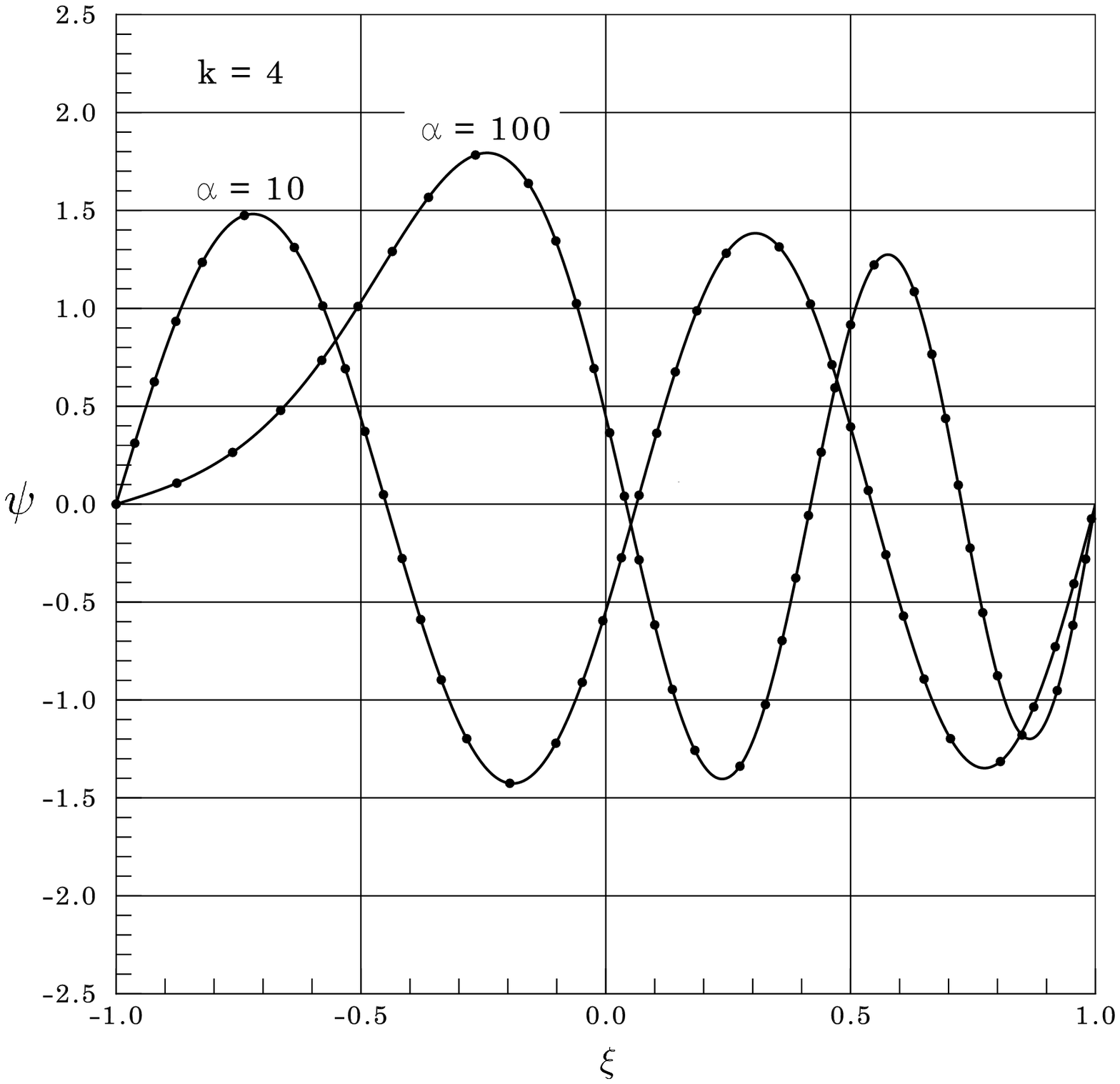}}     
   \caption{The first four eigenfunctions of an electric charge in uniform electrostatic field.\label{aef1-4.wef}}
\end{figure}

\section{External forces with the presence of the electric field\label{efeef}}

The availability of eigenfunctions in analytical form [\REq{evp.se.a.gs.2}] permits the straightforward determination of their derivative
\begin{equation}\label{aefd}
   \pd{}{\psi}{\eta} = \sqrt{\dfrac{2\alpha^{1/3}}{J}} \cdot f'(\eta)
\end{equation}
The prime on the right-hand side of \REq{aefd} indicates derivation with respect to $\eta$.
The eigenfunctions' derivative with respect to the old independent variable $\xi$ is obtained easily by taking advantage of the variable transformation enforced in \REq{ivt}
\begin{equation}\label{aefd.xi}
  \pd{}{\psi}{\xi} = - \alpha^{1/3} \pd{}{\psi}{\eta} = - \sqrt{\dfrac{2\alpha}{J}} \cdot f'(\eta)
\end{equation}
Therefore, the external forces at the left/right boundary turn out to be respectively
\begin{equation}\label{bef.c}
  \begin{cases}
     \left( \pd{}{\psi}{\xi} \pd{}{\psi^{\ast}}{\xi} \right)_{\xi=-1} = \dfrac{2\alpha}{J} \cdot f'(\etah)^{2} & \qquad \text{left boundary}  \\[3ex]
     \left( \pd{}{\psi}{\xi} \pd{}{\psi^{\ast}}{\xi} \right)_{\xi=+1} = \dfrac{2\alpha}{J} \cdot f'(\etab)^{2} & \qquad \text{right boundary}
  \end{cases}
\end{equation}
\begin{figure}[h]
   \hspace*{0.95\mathindent}
   \resizebox{\sfsize}{!}{\includegraphics*[trim=10 5 35 60]{\figdir/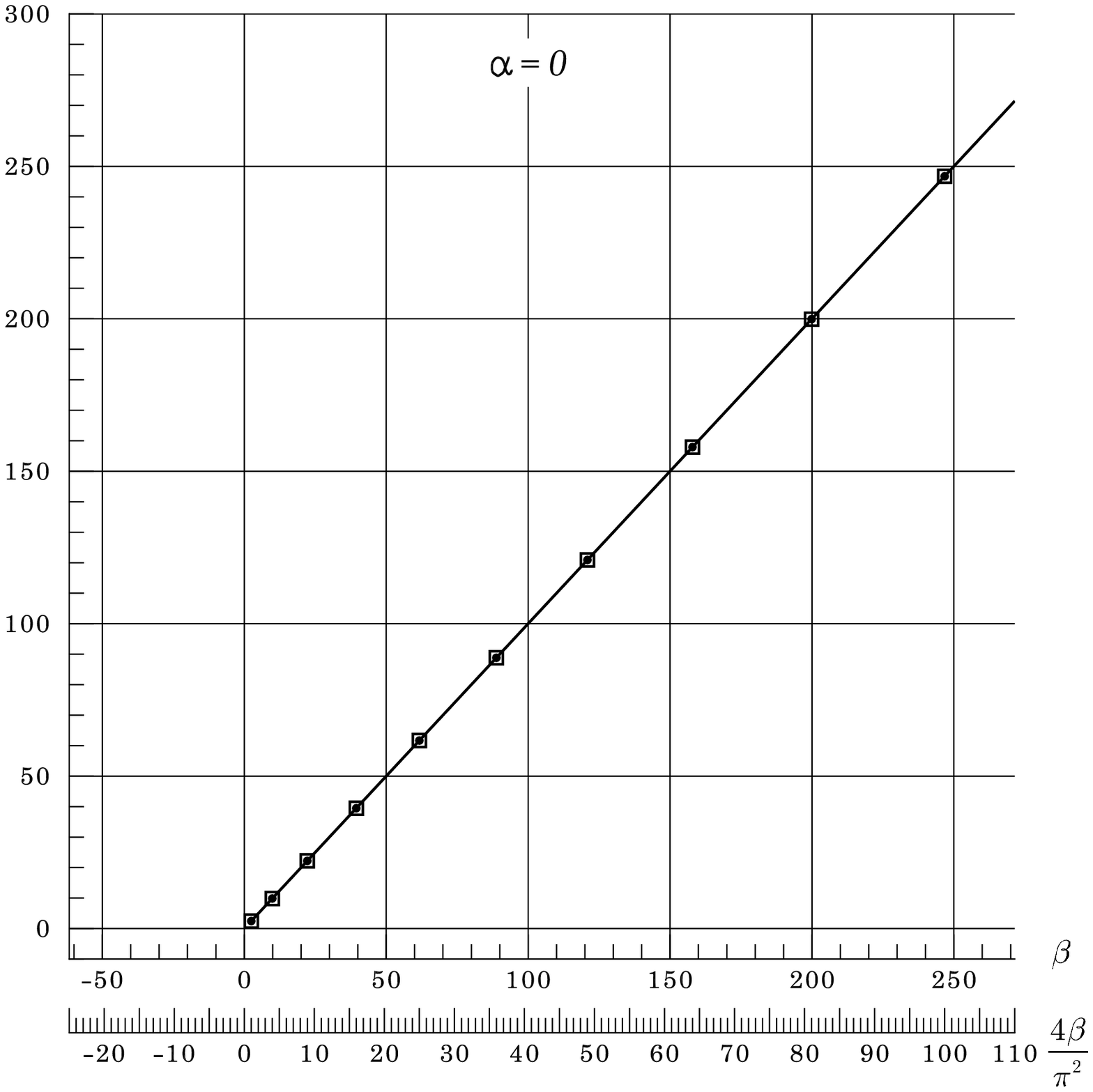}} 
   \resizebox{\sfsize}{!}{\includegraphics*[trim=10 5 35 60]{\figdir/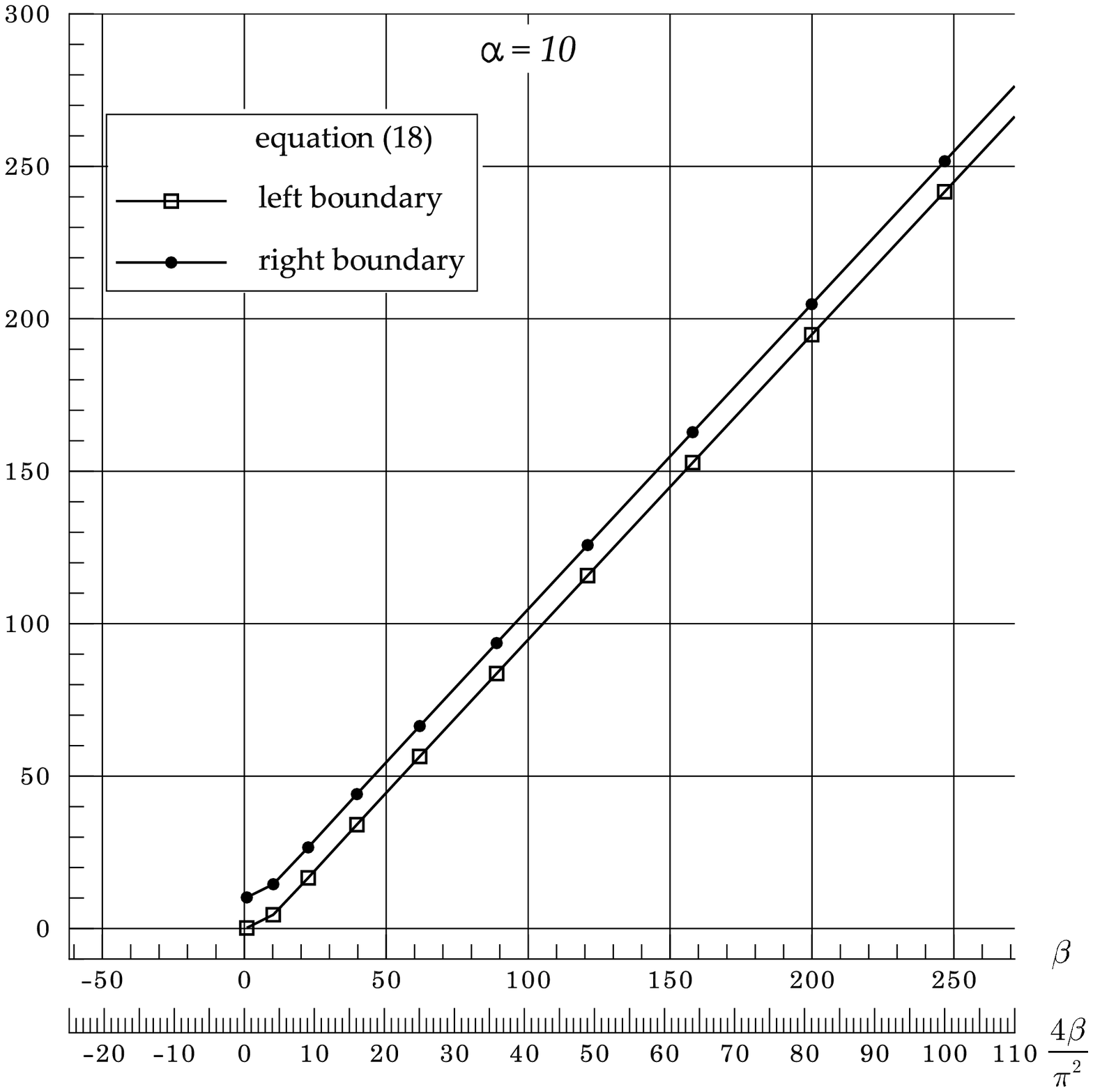}} \\[2ex] 
   \hspace*{0.95\mathindent}
   \resizebox{\sfsize}{!}{\includegraphics*[trim=10 5 35 60]{\figdir/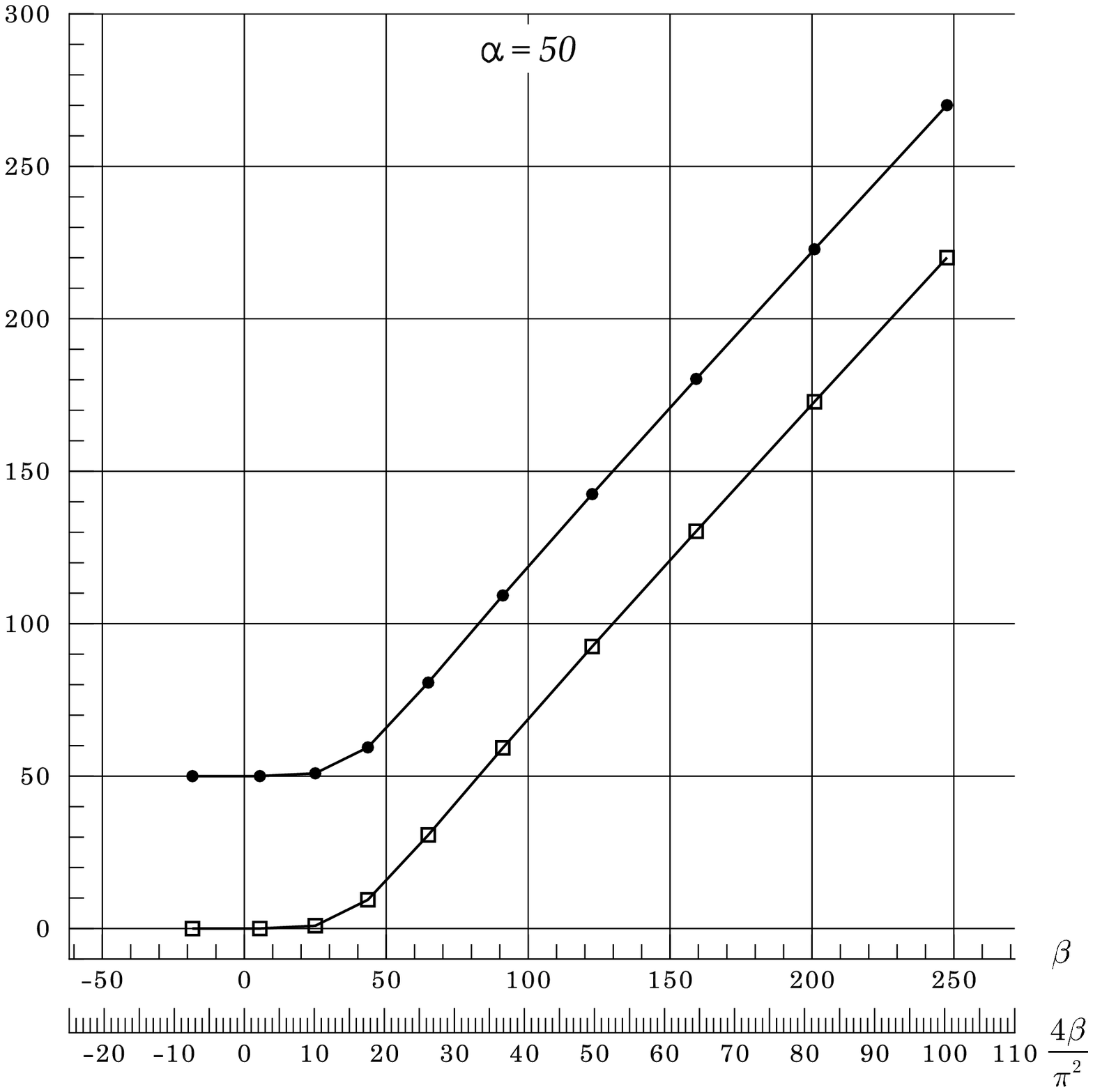}} 
   \resizebox{\sfsize}{!}{\includegraphics*[trim=10 5 35 60]{\figdir/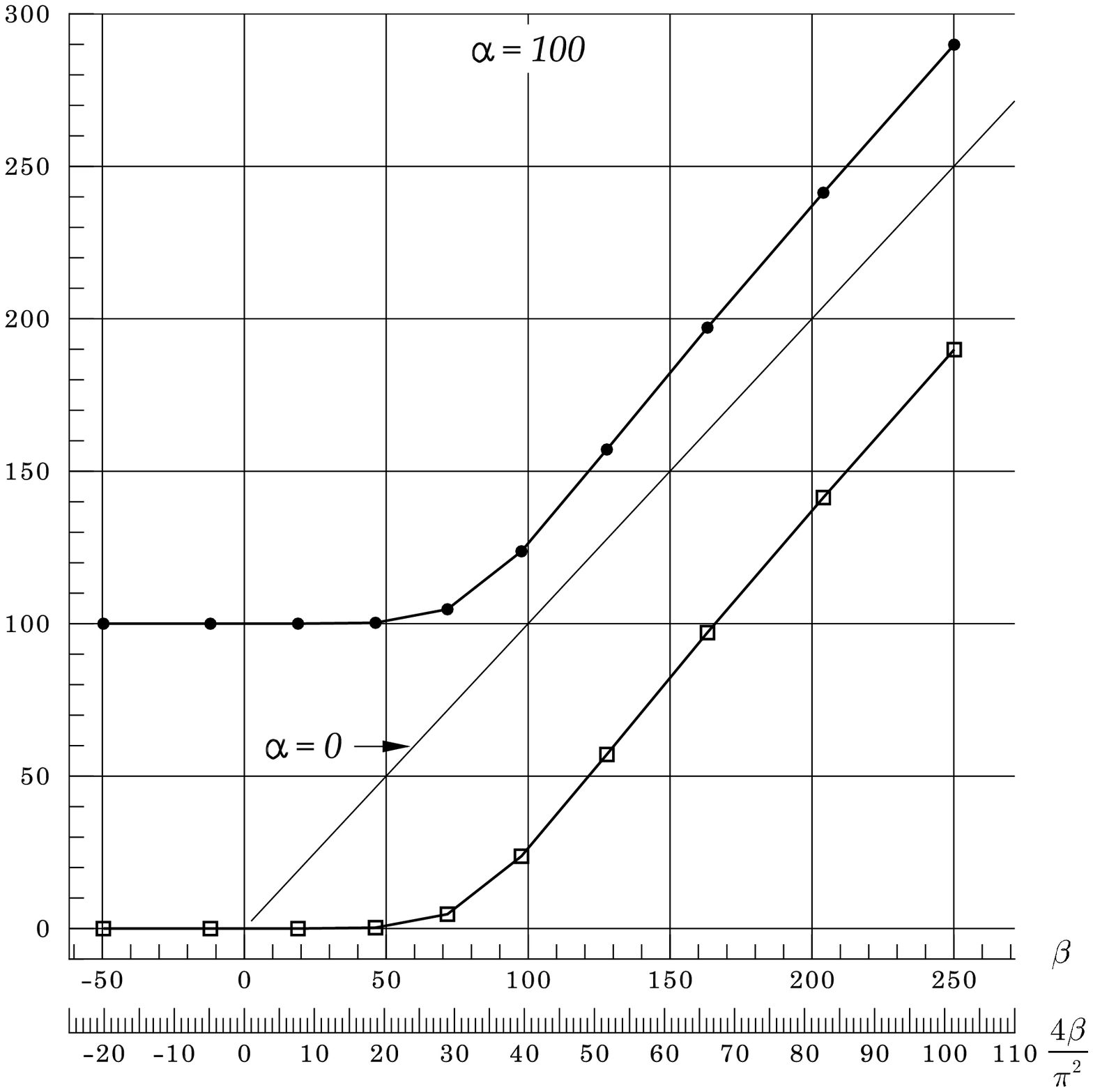}}     
   \caption{The (halved) external forces at the boundaries for the first ten eigenstates with increasing electric field.\label{bf.wef}}
\end{figure}
We have plotted them, divided by 2 for convenience, in \Rfi{bf.wef} {\authorcorr for the first ten eigenstates with increasing electric field}.
The top-left graph corresponds to the absence of electric field. 
The external forces are equal and vary linearly in compliance with \REq{ef.dv.c}; the projections of the symbols onto the axis $4\beta/\pi^{2}$ follow the sequence $k^{2}$ in compliance with \Reqma{99}.
The effect of the electric field is evident from the top-right and bottom graphs.
The external forces differ because they must equilibrate the electrical force.
Beginning from the lower eigenstates, the electric field destroys the linearity and makes the external forces basically independent from the energy eigenvalues.
The bottom graphs indicate that the number of affected eigenstates increases with the electric-field intensity: there are three for \mbox{$\alpha=50$} and four for \mbox{$\alpha=100$}.
For these eigenstates, the left-boundary external force (hollow squares) vanishes and the right-boundary external force (solid circles) is the only one that equilibrates the electric force.
For any specified electric-field intensity, there is always a transition from constant to linear dependence but the linearity is different from the one existing without electric field, not only, as expected, in magnitude but also in slope; in order to put in evidence the latter characteristic, we have echoed the trace of the case \mbox{$\alpha=0$} from the top-left graph onto the bottom-right graph corresponding to the case \mbox{$\alpha=100$}.
The graphs offer also the evidence that the difference between the external forces keeps constant and coincides with the value of $\alpha$; this is the graphical verification of \Reqma{105}, perfectly aligned to the numerical one we gave in \cite{dg2021ejp}.
{\refereeone The physical interpretation of the situation portrayed in \Rfi{bf.wef} is that, as far as the boundary forces are concerned, the effect of the electric field is never negligible and is felt in all the eigenstates, no matter how high-lying they are.
This result is in flagrant contraposition with the intuitive expectation that the effect of the electric field should be negligible when \mbox{$k\rightarrow\infty$} because, in that case, \mbox{$\alpha|\xi|\leq\alpha\ll\beta$} and, consequently, the electric-field term in \REq{evp.se} would be looked at as irrelevant with respect to the right-hand side.
\begin{table}[h] \refereeone
   \hspace*{\mathindent}
   \resizebox{.5\ftwidth}{!}{\includegraphics*{\figdir/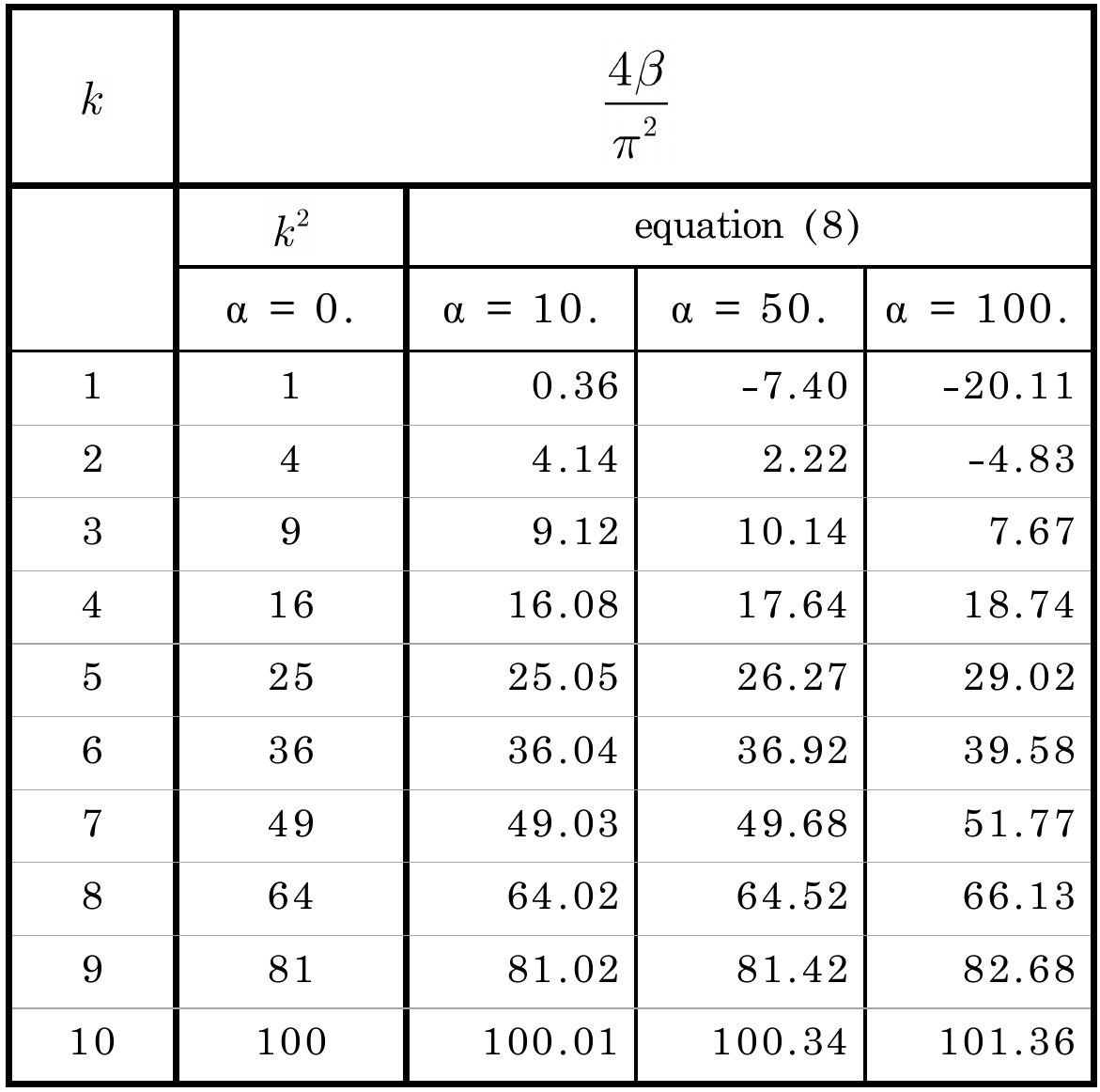}}  
   \caption{Scaled eigenvalues calculated from \Reqma{99} and   \REq{evg}\label{tables}}
\end{table}
This occurrence is true for the eigenvalues, as evidenced by \Rta{tables}, but not for the boundary forces: they are always affected by the electric field, as proved by the graphs of \Rfi{bf.wef}, inasmuch as they must balance the same electric force in each eigenstate [\Reqma{105}].}

\section{Analytical proof of \Reqma{105}\label{ap}}
The presence of $\alpha$ on the right-hand sides of \REqq{bef.c} caught particularly our attention with regard to \Reqma{105} that endorses the equilibrium between the electrical force and the external forces and that we verified numerically, as shown in the 2nd column from right in Table 1 of \cite{dg2021ejp}, and graphically in \Rfi{bf.wef}.
The substitution of \REqq{bef.c} into \Reqma{105} gives
\begin{equation}\label{efe}
   \alpha = \frac{1}{2} \left[ \dfrac{2\alpha}{J} \cdot f'(\etab)^{2} - \dfrac{2\alpha}{J} \cdot f'(\etah)^{2} \right]
\end{equation}
that{\authorcorr, after simplification, reduces} to
\begin{equation}\label{efe.s}
   J = f'(\etab)^{2} - f'(\etah)^{2}
\end{equation}
\REqb{efe.s} provides an expression of the integral $J$ in terms of the first derivatives of the function $f(\eta)$ evaluated at the boundaries and is the analytical shortcut, mentioned in \Rse{asev}, that we have systematically used in our calculations.
However, according to the way we have obtained it, \REq{efe.s} should be read as an expectation required by the force equilibrium established by \Reqma{105} but {\authorcorr cannot be considered} a rigorous {\authorcorr mathematical} proof.

The explicit mathematical proof of \REq{efe.s}, and, by reflection, of \Reqma{105}, exists but it requires to follow another path.
A few preliminary ingredients are necessary for that purpose.
The function $f(\eta)$ is a linear combination of the Airy functions [\REq{feta}] and, {\authorcorr as such}, it is also a solution of the Airy differential equation
\begin{subequations}\label{ing}
	\begin{equation}\label{ing1}
	   f''-\eta\,f = 0
	\end{equation}
    \addtocounter{footnote}{1}%
	which complies with the boundary conditions\addtocounter{footnote}{1}\,\footnote{Check \REq{evp.se.a.gs.2} evaluated at the boundaries, with due account of \REq{w}.}
	\begin{equation}\label{ing.bc}
	   f(\etah) = f(\etab) = 0
	\end{equation}
	Further differentiation of \REq{ing1} gives
	\begin{equation}\label{ing2}
	   f''' - f - \eta\,f' = 0
	\end{equation}
	which can be conveniently rearranged as
	\begin{equation}\label{ing3}
	   \dfrac{f}{\eta} = \dfrac{f'''}{\eta}  - f' 
	\end{equation}
\end{subequations}
We can proceed now with the calculation of the integral [\REq{J}].
We expand the integrand by taking advantage of \REqqc{ing}{(a,c,d)}
\begin{subequations}\label{int}
   \begin{equation}\label{int.1}
     \begin{split}
        J & = \int_{\etab}^{\etah} f^{2} d\eta = \int_{\etab}^{\etah} f \cdot f \,d\eta = \int_{\etab}^{\etah} f \cdot \dfrac{f''}{\eta} \,d\eta 
            = \int_{\etab}^{\etah} \dfrac{f}{\eta} \cdot f'' \,d\eta 
          \\[1ex]  
          & = \int_{\etab}^{\etah} \left( \dfrac{f'''}{\eta}  - f' \right) \cdot f'' \,d\eta 
          \\[1ex]  
          & 
            = \int_{\etab}^{\etah} \dfrac{f'''}{\eta} \cdot f'' \,d\eta - \int_{\etab}^{\etah} f' \cdot f'' \,d\eta
          \\
     \end{split}
   \end{equation}
   The second integral on the bottom line of \REq{int.1} is easily calculated
   \begin{equation}\label{int.1.2}
      \int_{\etab}^{\etah} f' \cdot f'' \,d\eta = \int_{\etab}^{\etah} f' \cdot \,df' = \frac{1}{2} \left[ f'(\etah)^{2} - f'(\etab)^{2} \right]
   \end{equation}
   and returns an encouraging expression in view of reaching \REq{efe.s}.
   The first integral on the bottom line of \REq{int.1} requires further manipulation
   \begin{equation}\label{int.1.1}
      \int_{\etab}^{\etah} \dfrac{f'''}{\eta} \cdot f'' \,d\eta = \int_{\etab}^{\etah} \dfrac{f''}{\eta} \cdot f''' \,d\eta
      = \int_{\etab}^{\etah} f \cdot  \,df'' 
      = \underline{\int_{\etab}^{\etah} \,d \left(f \cdot  f''\right)} -  \int_{\etab}^{\etah} f'' \cdot f'\,d\eta
   \end{equation}
   The underlined integral in \REq{int.1.1} vanishes because $f$ and $f''$ themselves vanish at the boundaries as a consequence of the boundary conditions [\REq{ing.bc}].
   Therefore, \REq{int.1.1} reduces to
   \begin{equation}\label{int.1.1a}
	  \int_{\etab}^{\etah} \dfrac{f'''}{\eta} \cdot f'' \,d\eta = -  \int_{\etab}^{\etah} f'' \cdot f'\,d\eta 
   \end{equation}
   We can now substitute \REq{int.1.1a} into the bottom line of \REq{int.1}, take advantage of \REq{int.1.2} and obtain the proof we were looking for
   \begin{equation}\label{efe.s.app}
            J = f'(\etab)^{2} - f'(\etah)^{2}
   \end{equation}
   \REqb{efe.s.app} also constitutes the solid analytical proof of \Reqma{105}\,.
\end{subequations}

\section*{Acknowledgments}
We are grateful to K. Moulopoulos for his interest in and feedback about our main article. 
The additional study described in this brief addendum has been motivated by his inspiring question quoted in \Rse{intro}.


\vspace*{.5\baselineskip}
\section*{References}
\bibliographystyle{iopart-num}
     

\bibliography{/Users/dg/Library/texmf/bibtex/bib/mybibreflibrary.bib}

\end{document}